\documentclass[10pt]{article}
\pdfoutput=1
\usepackage{jheppub,amsmath,amssymb}
\usepackage{enumerate}
\usepackage{bm}
\usepackage{bbm}
\usepackage{enumitem}
\usepackage[usenames,dvipsnames]{xcolor}
\usepackage{amsmath}
\usepackage{amsfonts}
\usepackage{amssymb}
\usepackage{graphicx}
\usepackage{hhline}
\usepackage{subfig}
\usepackage{hyperref}
\usepackage{color}
\usepackage{verbatim}
\usepackage{amsmath,amsthm,amssymb}
\usepackage{lscape}
\usepackage{float}
\usepackage{caption}
\usepackage{lscape}
\usepackage{breqn}
\usepackage{multicol}
\usepackage{graphics}
\usepackage{tikz,todonotes}
\usepackage[utf8]{inputenc}
\usepackage{autobreak}
\usepackage{verbatim}
\usetikzlibrary{arrows,positioning,decorations.markings,decorations.pathmorphing,calc}
\pgfdeclarelayer{edgelayer}
\pgfdeclarelayer{nodelayer}
\usetikzlibrary{decorations.pathreplacing}
\pgfsetlayers{edgelayer,nodelayer,main}
\tikzset{none/.style={draw=none}}
\tikzset{new edge style 2/.style={black}}
\tikzset{new style 0/.style={black}}
\tikzset{rednode/.style={draw=none, scale=0.3pt,fill=red,circle, draw}}
\tikzset{redline/.style={line width=0.3mm,red}}
\tikzset{greyE/.style={line width=0.1mm,gray}}
\usepackage[utf8]{inputenc}
\usetikzlibrary{arrows,positioning,decorations.markings,decorations.pathmorphing,calc}

\definecolor{hyperref}{RGB}{026,028,087}

\newcommand{\beq}{\begin{equation}}
\newcommand{\eeq}{\end{equation}}
\newcommand{\bea}{\begin{eqnarray}}
\newcommand{\eea}{\end{eqnarray}}
\def\be{\begin{equation}}
\def\ee{\end{equation}}

\def\beq{\begin{equation}}
\def\eeq{\end{equation}}

\newcommand{\M}{M}

\renewcommand{\L}{\mathcal L}

\def\be{\begin{equation}}
\def\ee{\end{equation}}
\def\ba{\begin{eqnarray}}
\def\ea{\end{eqnarray}}

\def\nn{\nonumber}

\def\d{\mathrm{d}}

\usepackage[normalem]{ulem}

\def\ba{\begin{eqnarray}}
\def\ea{\end{eqnarray}}

\def\L{\mathcal{L}}

\def\stu{St\"uckelberg }
\newcommand{\Lic}{{Lichnerowicz }}

\def\d{\mathrm{d}}
\def\mn{_{\mu \nu}}
\def\ab{_{\alpha \beta}}

\def\({\left(}
\def\){\right)}

\def\ie{{\em i.e. }}

\begin{document}

\title{Positivity Constraints on Interacting Pseudo-Linear Spin-2 Fields}

\author[a]{Lasma Alberte,}
\author[a,b]{Claudia de Rham,}
\author[a]{Arshia Momeni,}
\author[a]{Justinas Rumbutis,}
\author[a,b]{Andrew J. Tolley}
\affiliation[a]{Theoretical Physics, Blackett Laboratory, Imperial College, London, SW7 2AZ, U.K.}
\affiliation[b]{CERCA, Department of Physics, Case Western Reserve University, 10900 Euclid Ave, Cleveland, OH 44106, USA}

\emailAdd{l.alberte@imperial.ac.uk}
\emailAdd{c.de-rham@imperial.ac.uk}
\emailAdd{arshia.momeni17@imperial.ac.uk}
\emailAdd{j.rumbutis18@imperial.ac.uk}
\emailAdd{a.tolley@imperial.ac.uk}

\abstract{We explore the effective field theory for single and multiple interacting pseudo-linear spin-2 fields. By applying forward limit positivity bounds, we show that among the parameters contributing to elastic tree level scattering amplitude, there is no region of compatibility of the leading interactions with a standard local UV completion. Our result generalizes to any number of interacting pseudo-linear spin-2 fields. These results have significant implications for the organization of the effective field theory expansion for pseudo-linear fields.}

\maketitle


\section{Introduction}

The effective field theory description of massive spin fields, and in particular those of spin-2, is of interest for many reasons. Massive spin-2 states have been considered in the context of particle physics models \cite{Chivukula:2017fth,Bernal:2018qlk,Marzola:2017lbt,Babichev:2016bxi}, and play a central role in massive theories of gravity \cite{deRham:2014zqa}. Massive spin-2 states clearly play an important role in low energy descriptions of Kaluza-Klein and other braneworld constructions \cite{Overduin:1998pn,deRham:2014zqa,Bonifacio:2019ioc}. They also arise in the condensed matter context, for example as effective descriptions of the gapped collective excitation in fractional quantum Hall systems \cite{Gromov:2017qeb}. \\

In the relativistic context, it is natural to interpret all theories of massive spin particles in terms of a breaking of the symmetries of a massless particle. The central reason being that in a Lorentz invariant theory, at energies much higher than the mass of the particle, the states of a massive spin will naturally decompose into those of massless helicity modes. By virtue of their canonical normalization, the interaction scales for the different helicity states are different, and this has a significant impact on the organization of the low energy effective field theory expansion. The case of massive spin-2 has been well studied \cite{ArkaniHamed:2002sp,deRham:2014zqa}. For generic higher spins, the symmetry structures that give rise to interacting massless theories are necessarily infinite as there are strong theorems precluding interactions of finitely many spins with $s > 2$ \cite{Vasiliev:1995dn,Rahman:2015pzl,Rahman:2013sta}. The spin-2 case is special in that we already know of one description of infinitely many interacting massive spin-2 fields with one massless, namely Kaluza-Klein theory. The would-be infinite number of 4 dimensional diffeomorphism symmetries that would arise for decoupled massless gravitons, combine together to make a theory which respects a higher dimensional diffeomorphism symmetry. Rewritten in four dimensional terms, the higher dimensional diffeomorphism symmetry appears as a Kac-Moody type algebra \cite{Dolan:1983aa}. Hence, stated differently, Kaluza-Klein theory may be interpreted as a spontaneously broken version of a four dimensional theory with a Kac-Moody symmetry \cite{Dolan:1983aa}. \\

These infinite dimensional symmetry groups generically enforce an infinite number of closely spaced states \cite{Bonifacio:2019ioc} and so are not useful descriptions in situations where there may exist a gap, such that there is a low energy effective description with only a finite number of spin-2 states. Such situations do occur in the condensed matter context \cite{Gromov:2017qeb} and it is interesting to explore this possibility in the Lorentz invariant context.  For a finite number of spin-2 and lower spin fields, the broken symmetry group will be finite dimensional. It is thus natural to ask, how many nonlinear extensions do there exist for the symmetries of set of free spin-2 fields. A known nonlinear symmetry, determines the symmetry breaking mechanism which in turn organizes the construction of the low energy effective theory. In the case of a single massive spin-2, which at free massless level has a copy of linear diffeomorphisms (spin-2 gauge invariance), it is known that there are only two nonlinear completions of the symmetry itself  \cite{Wald:1986bj}. Full diffeomorphisms, and the same linear spin-2 gauge invariance. Thus an interacting effective field theory of a single massive spin-2 field can either arise from a spontaneously broken diffeomorphism symmetry, or a spontaneous breaking of spin-2 gauge invariance  \cite{Wald:1986bj}. The former case corresponds to massive gravity and multi-gravity theories, and is by far the most commonly assumed scenario. The latter is sometimes referred to as pseudo-linear spin-2 massive gravity \cite{Hinterbichler:2013eza}, however we just refer to these as pseudo-linear spin-2 fields as they have no connection with gravity per se. Pseudo-linear spin-2 fields could prove to be a useful description for the EFT of excited spin-2 mesons, similar to those that arise in the condensed matter context where they have no immediate connection with gravitational physics. \\

Whether or not any single or multiple interacting massive spin-2 fields could have a standard UV completion remains as yet unclear \cite{deRham:2018dqm}. By assuming the UV completion to be  Lorentz invariant, causal and unitary it is possible to derive particular bounds on the scattering amplitudes of the low energy EFTs, so called positivity bounds, which restrict the allowed parameter space of the EFT \cite{Pham:1985cr,Ananthanarayan:1994hf,Adams:2006sv,Cheung:2016yqr,
Bonifacio:2016wcb,Bellazzini:2016xrt,deRham:2017xox,
Bellazzini:2017fep,deRham:2018qqo,alberte2020positivity}. We stress that a failure to satisfy these bounds does not imply the EFT is inconsistent but only that it could not have a standard UV completion, but this does not preclude a non-standard UV completion \cite{deRham:2017xox}. Indeed one of the most likely standard assumptions to fail is that of locality, since it is not expected that a gravitational theory respects polynomial (or exponential) boundedness \cite{Giddings:2009gj}, and this is a central assumption in the derivation of positivity bounds \cite{Keltner:2015xda}. Remarkably two dimensional ghost-free massive gravity does admit a known UV completion, since when coupled to a conformal field theory with central charge $c=24$ it is equivalent to the worldsheet theory of a critical string \cite{Tolley:2019nmm}, and more generically that of a non-critical string. The latter are known to have a worldsheet S-matrix which violates polynomial/exponential boundedness and so does not respect standard locality requirements \cite{Dubovsky:2012wk}.\\

In a recent work \cite{alberte2020positivity} we applied the positivity bounds to EFTs of interacting multiple spin-2 fields described in \cite{alberte2020eft}, which were extensions of ghost-free massive gravity. In this paper we apply these bounds to effective field theories of one or more interacting pseudo-linear massive spin-2 fields. The particular case of a single pseudo-linear spin-2 field was considered already in \cite{Bonifacio:2016wcb} where it was shown that for the leading `ghost-free' interactions some of the definite helicity positivity bounds have to be  marginal, unlike its close cousin `ghost-free massive gravity' \cite{Cheung:2016yqr,deRham:2017xox,deRham:2018qqo,alberte2020positivity}. Another interesting example where positivity bounds are marginal is in the case of massless Galileons \cite{Nicolis:2008in}. In this theory the 2-2 tree level scattering amplitude grows as the third power of the Mandelstam variables which means that the positivity bounds are marginal \cite{Adams:2006sv}. Interestingly, massless Galileons arise as the  decoupling limit of massive gravity \cite{deRham:2010ik}, which would seem to suggest that massive gravity cannot have a standard UV completion. However, it was shown in \cite{Cheung:2016yqr} that there exists a compact region in the parameter space for which the scattering amplitudes of this theory are compatible with forward limit positivity bounds and this has been further considered in \cite{deRham:2017xox,Bellazzini:2017fep,deRham:2018qqo,alberte2020positivity}. This is because taking a decoupling limit is distinct to considering the low-energy EFT, and while Galileons are massless in the decoupling limit of massive gravity (where the mass is sent to zero), those modes remain massive in the low-energy limit of massive gravity. As shown in \cite{deRham:2017imi} this distinction is crucial in satisfying the positivity bounds. \\

The rest of this paper is organized as follows: in Section~\ref{sec:main} we review the EFT for a single and for multiple interacting massive pseudo-linear spin-2 fields.  We then discuss the positivity bounds in Section~\ref{pos bounds} before explaining how they constrain the EFT in Section~\ref{sec:bounds} for one or two interacting pseudo-linear spin-2 fields. We revisit the single spin-2 case considered in \cite{Bonifacio:2016wcb} and find and close a loophole in the argument that it is ruled out. We then extend our result to an arbitrary number of massive pseudo-linear spin-2 fields in section~\ref{gen}.  We end with some outlooks and discussions in Section~\ref{sec:conclusions}.
Details on the polarization structure are given in Appendix~\ref{sec:transv} and the indefinite bounds for the higher order operators are given in Appendix~\ref{app:indef}.

\section{EFT of Interacting Pseudo Linear Spin-2 Fields}\label{sec:main}

\subsection{Single field}
We start by considering the standard Fierz--Pauli linear Lagrangian for a single massive spin-2 field \cite{Fierz:1939ix}
\ba
\L_{\rm FP}=- h^{\mu \nu }\mathcal{E}^{\alpha\beta}\mn h\ab-\frac 12 m_1^2 \([h^2]-[h]^2\)\,,
\ea
where the squared brackets denote the trace with respect to the Minkowski metric and $\mathcal E$ stands for the \Lic operator defined as
\be\label{quadratic}
\mathcal E^{\alpha\beta}_{\mu\nu}h_{\alpha\beta}=-\frac{1}{2}\left[\Box h_{\mu\nu}-\partial_\alpha\partial_{\mu}h^\alpha_{\nu}-\partial_\alpha\partial_\nu h^\alpha_\mu+\partial_\mu\partial_\nu h-\eta_{\mu\nu}\left(\Box h-\partial_\alpha\partial_\beta h^{\alpha\beta}\right)\right]\,.
\ee
While the Fierz-Pauli action breaks linear diffeomorphisms, it is ghost-free and propagates five degrees of freedom in four-dimensions. The breaking of linearized diffeomorphism can be `restored' by the introduction of four linear \stu fields, three of which are dynamical and describe the propagating states of the helicity-1 and helicity-0 modes. \\

When supplementing the Fierz-Pauli action with non-linear interactions, one possibility is to promote the linearized diffeomorphism (spin-2 gauge invariance) symmetry to full nonlinear diffeomorphisms (general coordinate transformations) and the resulting theory is then closely linked to a gravitational theory. Alternatively, and this will be the approach considered here,  one can view the spin-2 states' masses arising from the breaking of a linearized diffeomorphism symmetry, even at the interacting level, and therefore maintain the \stu fields as introduced linearly (see \cite{deRham:2011qq} for a discussion on the distinction between the non-linear \stu and the linear or helicity approach). Such fields are referred to as pseudo-linear spin-2 fields \cite{Hinterbichler:2013eza}. The pseudo-linear reflects the fact that the gauge symmetry is linear, even though we consider interactions. \\

It was established in \cite{Hinterbichler:2013eza} that only three pseudo-linear ghost-free terms could be added to the Fierz-Pauli action in four dimensions. Those were found by requiring that: $(i)$ they preserve the number of degrees of freedom of the linear Fierz-Pauli theory; $(ii)$ when possible they arise as terms leading order in perturbations of a non-linear field that satisfies $(i)$. Two of these terms arise straightforwardly as decoupling limits of the standard ghost-free interactions for massive gravity  \cite{deRham:2010kj}. Indeed on taking the non-linear fields $\mathcal K^\mu_\nu=\delta^\mu_\nu-\left(\sqrt{g^{-1}f}\right)^\mu_\nu$ and expressing the metric as $g_{\mu\nu}=\eta_{\mu\nu}+\frac{h_{\mu\nu}}{M_1}$, where $M_1$ is the scale of non-linearities, then in the decoupling limit $M_1 \rightarrow \infty$ with $\alpha_3$ and $\alpha_4$ scaled appropriately, the usual ghost-free massive gravity interactions become
\be\label{pseudo_pot}
\begin{split}
&\mathcal L_{0,3}=\varepsilon\varepsilon Ihhh= [h]^3-3[h][h^2]+2[h^3]\;,\\
&\mathcal L_{0,4}=\varepsilon\varepsilon hhhh=[h]^4-6[ h^2][ h]^2+8[h^3][h]+3[h^2]^2-6[h^4]\;.
\end{split}
\ee
Here and henceforth we use the shorthand notation
\begin{equation}
\label{eq:L034}
\begin{split}
&\varepsilon\varepsilon I^{4-n}\mathbb X^n\equiv\varepsilon_{\mu_1\dots\mu_n\mu_{n+1}\dots\mu_4}\varepsilon^{\nu_1\dots\nu_n\nu_{n+1}\dots\nu_4}\mathbb X^{\mu_1}_{\nu_1}\dots\mathbb X^{\mu_n}_{\nu_n}\delta^{\mu_{n+1}}_{\nu_{n+1}}\dots\delta^{\mu_4}_{\nu_4}\,,\\
&\varepsilon\varepsilon I^{4-n} \partial^2\mathbb X^n\equiv\varepsilon_{\mu_1\dots\mu_n\mu_{n+1}\dots\mu_4}\varepsilon^{\nu_1\dots\nu_n\nu_{n+1}\dots\nu_4} (\partial^{\mu_1} \partial_{\nu_1}\mathbb X^{\mu_2}_{\nu_2})\dots\mathbb X^{\mu_n}_{\nu_n}\delta^{\mu_{n+1}}_{\nu_{n+1}}\dots\delta^{\mu_4}_{\nu_4}\,,
\end{split}
\end{equation}
where $[\mathbb X]$ denotes the trace of the matrix/tensor $\mathbb X$. There are no ghost-free terms with four or more derivatives in four spacetime dimensions (closely related to the Lovelock theorem \cite{Lanczos:1938sf,Lovelock:1971yv}). However as clarified in \cite{Hinterbichler:2013eza}, the pseudo-linear theory allows a two-derivative ghost-free cubic interaction term, first found in \cite{Folkerts:2011ev}, and is given by
\be\label{pseudo_kin}
\mathcal L_{2,3}=\varepsilon\varepsilon (\partial^2h)hh\,.
\ee
This term is not the Einstein-Hilbert term cubic interaction since the latter would only be ghost-free if we consider full non-linear diffeomorphism invariance at the non-linear level. Indeed this term appears to be an isolated feature of the pseudo-linear theory and does not have an equivalent ghost-free structure when the symmetry is nonlinear diffeomorphism \cite{deRham:2013tfa,deRham:2015rxa,deRham:2015cha,Matas:2015qxa}, {\ie} in the massive gravity context.\\

In what follows,  we shall not be concerned with the theory defined uniquely by the ghost-free interactions, but rather by the effective field theory with the highest cutoff, similarly to the logic followed in \cite{alberte2020eft,alberte2020positivity}. From this perspective the interactions \eqref{pseudo_pot} are to be regarded as the leading interactions in a Wilsonian effective action which contains an infinite number of terms. Thus the leading terms used to describe the effective action of a single pseudo-linear spin-2 field are taken to be
\be\label{action_single}
g_*^2 \mathcal L=\mathcal L_{\rm FP}+\frac{a_1}{2M_1}\mathcal L_{2,3}+\frac{m_1^2\kappa_3^{(h)}}{4M_1} \mathcal L_{0,3}+\frac{m_1^2\kappa_4^{(h)}}{4M_1^2}\mathcal L_{0,4}\,,
\ee
where $a_1$ and $\kappa_{3,4}^{(h)}$ are dimensionless coupling constants. The choice of scales for the coefficients in this leading effective action will be defined in analogy to the general case \cite{alberte2020eft,alberte2020positivity} by identifying a spin-2 interaction scale $M_1$ and organizing the pseudo-linear theory as an EFT with the strong coupling scale $\Lambda_3=(m_1^2M_1)^{1/3}$. In addition, as in  \cite{alberte2020eft,alberte2020positivity} we have include an overall weak coupling parameter $g_*$ which conveniently suppresses loops if $g_* \ll 1$ allowing us to apply tree level positivity bounds only.

In writing \eqref{action_single}, it is worth remembering that if the interactions $\L_{0,3/4}$ were generic (not given by \eqref{eq:L034}), they would lead to a  ghost hidden in the higher derivative interactions of the helicity-0 mode of the massive spin-2 generically appearing at the scale $\Lambda_5=(m_1^4M_1)^{1/5}$ \cite{ArkaniHamed:2002sp,Deffayet:2005ys}. These dangerous higher derivative interactions can only be avoided by the special tunings in the non-derivative interaction terms corresponding to setting $\L_{0,3/4}$ to their expression given in \eqref{eq:L034}. This was  first recognized in \cite{deRham:2010kj} and applied to the pseudo-linear case in \cite{Hinterbichler:2013eza}. Since these tunings lead to a theory with a higher cutoff scale they are technically natural. Indeed, after replacing the metric with $h_{\mu\nu}\to \frac{\partial_\mu\partial_\nu\pi}{m_1^2}$ one observes that all the higher derivative terms drop out because of the double-epsilon structure in both derivative and non-derivative interactions \eqref{pseudo_pot} and \eqref{pseudo_kin}. The surviving subleading terms correspond to interactions which arise at the $\Lambda_3$ scale. Indeed the decoupling limit of the pseudo-linear theory is identical in form to the general case, namely it looks like a massless spin-2 coupled to a Galileon and a Maxwell field. Thus \eqref{action_single} defines the leading interactions of what we mean by a $\Lambda_3$ theory of a single pseudo-linear spin-2 field.

\subsection{Two fields}
In this work we are mainly interested in the theory of two or more interacting pseudo-linear massive spin-2 fields \cite{Bonifacio:2019pfg}. For now, we shall focus on two interacting pseudo-linear spin-2 fields which can easily be generalized to an arbitrary number of pseudo-linear fields in section~\ref{gen}. In four dimensions each of the two spin-2 can be separately described by the pseudo-linear action \eqref{action_single} given above. We can then couple the two spin-2 fields  as follows
\be\label{model0}
\begin{split}
 g_*^2 \L=&\,\L_{\rm FP 1}+\frac{a_1}{2M_1}\varepsilon\varepsilon (\partial^2h)hh+\frac{m_1^2\kappa_3^{(h)}}{4M_1}\varepsilon\varepsilon Ihhh+\frac{m_1^2\kappa_4^{(h)}}{4M_1^2}\varepsilon\varepsilon hhhh\\
+&\L_{\rm FP 2}+\frac{a_2}{2M_2}\varepsilon\varepsilon (\partial^2f)ff+\frac{m_2^2\kappa_3^{(f)}}{4M_2}\varepsilon\varepsilon Ifff+\frac{m_2^2\kappa_4^{(f)}}{4M_2^2}\varepsilon\varepsilon ffff\;\\
+&\frac{a_3}{2M_1}\varepsilon\varepsilon (\partial^2h)hf+\frac{a_4}{2M_2}\varepsilon\varepsilon (\partial^2 h)ff+\frac{a_5}{2M_2}\varepsilon\varepsilon (\partial^2 f)fh+\frac{a_6}{2M_1}\varepsilon\varepsilon (\partial^2 f)hh\;\\
+&\frac{m_2^2}{4}\left[\frac{2c_1}{M_1}\varepsilon\varepsilon I hhf+\frac{2c_2}{M_2}\varepsilon\varepsilon I hff+\frac{\lambda}{M_1M_2} \varepsilon\varepsilon hhff+\frac{d_{1}}{M_1^2} \varepsilon\varepsilon hhhf+\frac{d_{2}}{M_2^2} \varepsilon\varepsilon hfff\right]\,,
\end{split}
\ee
where the first two lines is the sum of the individual pseudo-linear EFTs for the decoupled fields, the third line gives the cubic two-derivative interactions between the two fields and the last line are the non-derivative interactions. On the last line we have also included the $d_1,d_2$ interactions which however do not contribute to elastic tree level scattering processes. They will therefore remain unconstrained by the positivity bounds explored in this work. The mass scales $m_1, M_1$ and $m_2,M_2$ are the mass and non-linearity scales of $h_{\mu\nu}$ and $f_{\mu\nu}$ respectively. We shall assume that there is no large hierarchy between the two sets of masses and parameterize their ratios as
\be\label{scaling0}
m_2\equiv m\,,\quad \frac{m_1}{m_2}\equiv x\,,\qquad M_2\equiv M\,,\quad \frac{M_1}{M_2}\equiv \gamma\,.
\ee
As in the single field case, \eqref{model0} should be regarded as the leading interactions in a Wilsonian effective action, organized with interactions at the $\Lambda_3$ scale, and this point should be remembered in interpreting the implications of the positivity bounds as we shall see.

\section{Positivity Bounds}\label{pos bounds}
In the following sections we shall apply the forward limit $2\to 2$ scattering amplitude positivity bounds to theories of interacting pseudo-linear spin-2 fields. The positivity bounds arise as certain conditions on the couplings in the low energy effective field theory due to the requirement of the existence of a local and unitary Lorentz invariant UV completion. Stated more formally, the knowledge of the analytic structure of the scattering amplitude in the complex $s$-plane allows one to relate a properly regulated contour integral $f$ of the low energy scattering amplitude to the total scattering cross section via the use of the optical theorem \cite{Pham:1985cr,Ananthanarayan:1994hf,Adams:2006sv}. In the framework of \cite{Cheung:2016yqr,deRham:2017xox,Bonifacio:2016wcb,Bellazzini:2016xrt,Bellazzini:2017fep,deRham:2018qqo,alberte2020positivity}, the positivity bounds can be imposed on the derivatives of the pole-subtracted forward limit ($t=0$) amplitude as:
\begin{equation}
\begin{split}\label{pos}
      f_{\lambda_1\lambda_2}=\frac{1}{2}\frac{\d^2}{\d s^2}(A^s_{\lambda_1\lambda_2\lambda_1\lambda_2}(s,0)-\text{poles})>0\,,
\end{split}
\end{equation}
where $\lambda_1,\lambda_2$ stand for the polarization states of the ingoing and outgoing particles which are assumed to be equal in elastic scattering. We refer the reader to \cite{deRham:2017zjm,alberte2020positivity} for derivation of \eqref{pos} in our current notations.

\subsection{Indefinite Scattering}
In the following Section we apply the positivity bounds \eqref{pos} on the elastic forward limit ($t=0$) two--to--two scattering amplitudes in the EFT given in \eqref{model0}. Our main focus will be the $hh\to hh$ (and the equivalent $ff\to ff$) scattering process allowing to rule out all the non-derivative self-couplings $\kappa_3^{(h)}$, $\kappa_4^{(h)}$, $\kappa_3^{(f)}$, $\kappa_4^{(f)}$, as well as all the cubic couplings $a_1,\dots,a_6$ and $c_1$, $c_2$. The only remaining quartic coupling sensitive to the tree-level positivity bounds, $\lambda$, is in turn ruled out by the $hf\to hf$ scattering, as in \cite{alberte2020positivity}. As in \cite{alberte2020positivity} we cannot exclude the interactions $d_1$ and $d_2$ since they do not contribute to elastic scattering at tree level.

We express the polarization states of the ingoing and outgoing particles in either the scalar-vector-tensor (SVT) or the transversity polarization basis, depending on convenience. While the former is the basis most commonly used in the context of scattering massive spin-2 particles (see, \emph{e.g.} \cite{Cheung:2016yqr}) we find that in some specific cases the transversity basis proves to be more useful \cite{deRham:2017zjm}. Also, while in many cases it turns out to be sufficient to only consider definite helicity states, we obtain the strongest constraints when considering arbitrary configurations of the helicity eigenstates. These indefinite polarization states of the ingoing and outgoing particles can be specified in the SVT basis as
\begin{align}
\label{indefpolhel}
\begin{split}
&\epsilon^{(1)}=\alpha_{T1}\epsilon_{T1}+\alpha_{T2}\epsilon_{T2}+\alpha_{V1}\epsilon_{V1}+\alpha_{V2}\epsilon_{V2}+\alpha_S\epsilon_{S}\,, \\
&\epsilon^{(2)}=\beta_{T1}\epsilon_{T1}+\beta_{T2}\epsilon_{T2}+\beta_{V1}\epsilon_{V1}+\beta_{V2}\epsilon_{V2}+\beta_S\epsilon_{S}\,,\\
&\epsilon^{(3)}=\epsilon^{(1)}\,,\\
&\epsilon^{(4)}=\epsilon^{(2)}\,,
\end{split}
\end{align}
where we have assumed that the polarizations of each ingoing and outgoing particle-pair (\emph{i.e.} of particles $1,3$ and $2,4$) are equal. In practice it proves sufficient in determining the strongest bounds to focus on real combinations, and so these ten real numbers $\alpha$, $\beta$ then entirely determine the configuration of helicities of the scattering process. The expressions of the polarization tensors, as well as the relation between the SVT and transversity basis are given in the Appendix~\ref{sec:transv}.

In the following subsection we briefly review and extend the positivity bound constraints on the single pseudo-linear spin-2 field EFT existing in the earlier literature~\cite{Bonifacio:2016wcb}.  We then impose the positivity bounds \eqref{pos} on the theory \eqref{model0} of two interacting pseudo-linear spin-2 fields in Section~\ref{sec:bounds}. We find that turning on interactions between the two fields forbids a standard UV completion for this EFT.

\subsection{Single Pseudo-Linear Spin-2 Field}

The positivity bounds for the case of a single pseudo-linear scalar field were previously studied in~\cite{Bonifacio:2016wcb} where it was argued that the theory was ruled out. In particular, keeping for now the same notation as in \cite{Bonifacio:2016wcb}, the  following bounds were obtained on the couplings of the EFT of a single pseudo-linear spin-2 field\footnote{Note that the $\lambda_i$'s in \eqref{pseudobounds} are the couplings considered in \cite{Bonifacio:2016wcb} and not polarization of the ingoing and outgoing states. The couplings are related to those in \eqref{action_single} as $\lambda_1=a_1$, $\lambda_3=\frac{3\kappa^{(h)}_3}{2}$, and $\lambda_4=6\kappa^{(h)}_4$, while $g_* M_p=M_1$. }
\begin{equation}
\label{pseudobounds}
    \begin{split}
         &f(TTTT)_{+}=\frac{9\lambda^2_{1}+4\lambda_{1}\lambda_{3}}{3m_1^2 M^2_{p}}>0\,,\\
        &f(TTTT)_{-}=\frac{\lambda^2_{1}}{m_1^2 M^2_{p}}>0\,,\\
        &f(TVTV)=-\frac{3\lambda^2_{1}+4\lambda_{1}\lambda_{3}}{16m_1^2 M^2_{p}}>0\,,\\
        &f(TSTS)=-\frac{4\lambda^2_{1}+2\lambda_{1}\lambda_{3}}{16m_1^2 M^2_{p}}>0\,,\\
        &f(VVVV)_{+}=-\frac{15\lambda^2_{1}+13\lambda_{1}\lambda_{3}+5\lambda^2_{3}}{12m_1^2 M^2_{p}}>0\,.
    \end{split}
\end{equation}
It was then argued that the fact that the second inequality imposes $\lambda_1\neq 0$ makes it impossible to satisfy the last inequality for any choice of values for $\lambda_1,\lambda_3$. In this sense the violation is marginal, since it is implicitly assumed that the bounds cannot be saturated, {\ie} that equality on the right hand side is not allowed. However if we allow the equality, then the rather trivial solution $\lambda_1=0$ would allow all but the $f(VVVV)_{+}$ bounds to be satisfied, with the latter enforcing $\lambda_3=0$. \\

There is however a small loophole in this argument, since the expressions on the left hand side of  \eqref{pseudobounds} are only the leading terms in the effective theory. As in the case of the massless Galileon which is ruled out marginally by positivity bounds \cite{Adams:2006sv}, a small correction to the effective field theory can allow these inequalities to be satisfied \cite{deRham:2017imi}. To be more specific, the meaning of the positivity bound $f>0$ is that $f$ has to be positive in the low energy effective field theory in question. However, any EFT contains an infinite number of operators. In particular, there are the leading order operators arising at the scale that sets the lowest interaction scale of the theory and there are the higher order operators arising, for instance, from loop corrections of heavy fields. These also contribute to the scattering processes and thus to $f$. Having this perspective in mind, when imposing positivity bounds on the leading order operators one should in fact only require
\be
f\gtrapprox 0\,.
\ee
The approximate equality in the above relation should mean that the positivity bounds are marginally satisfied and are sensitive to the higher order corrections. This possibility was also mentioned in \cite{Bonifacio:2016wcb}, however, it was argued that there are no higher derivative operators that can be added to the pseudo-linear theory without introducing additional degrees of freedom.  As was explained earlier, in the pseudo-linear theory the ghost-free operators $\L_{0,3},\L_{0,4},\L_{2,3}$ are the leading order operators arising at the $\Lambda_3$ scale and these are indeed the only ghost-free operators that can be written down in this theory. However, higher order terms do arise suppressed by the scale $\Lambda_3$. The  $\Lambda_3$-EFT was discussed for instance  in \cite{deRham:2018qqo,alberte2020eft} and the higher order operators take the form
\be
\mathcal L_{\rm h.o.}=\Lambda_3^4\,\tilde{\mathcal L}_{\rm h.o.}\left[\frac{\partial}{\Lambda_3},\frac{h}{M},\frac{m\partial A}{\Lambda_3^3},\frac{\partial^2\pi}{\Lambda_3^3}\right]\,.
\ee

To demonstrate explicitly how such higher order operators modify the positivity bounds let us consider an operator of the form
\begin{equation}\label{c_op}
\L_{\rm h.o.}=\frac{c}{4M^4}(\partial^{\rho}\partial^{\lambda}h^{\mu\nu})h_{\mu\nu}(\partial_{\rho}\partial_{\lambda}h^{\sigma \gamma})h_{\sigma \gamma}\,.
\end{equation}
The coupling $c$ will have the following contribution to the various bounds:
\begin{equation}
    \begin{split}
         &f(TTTT)_{+}=\frac{9\lambda^2_{1}+4\lambda_{1}\lambda_{3}}{3\Lambda^4_{2}}+\frac{c}{2M^4}>0\,,\\
        &f(TTTT)_{-}=\frac{\lambda^2_{1}}{\Lambda^4_{2}}+\frac{c}{4M^4}>0\,,\\
        &f(TVTV)=-\frac{3\lambda^2_{1}+4\lambda_{1}\lambda_{3}}{16\Lambda^4_{2}}+\frac{c}{4M^4}>0\,,\\
        &f(TSTS)=-\frac{4\lambda^2_{1}+2\lambda_{1}\lambda_{3}}{16\Lambda^4_{2}}+\frac{c}{4M^4}>0\,,\\
        &f(VVVV)_{+}=-\frac{15\lambda^2_{1}+13\lambda_{1}\lambda_{3}+5\lambda^2_{3}}{12\Lambda^4_{2}}+\frac{1}{M^4}\bigg(\frac{3s(s-4m^2)}{8m^4}+2\bigg) c>0\,.
    \end{split}
\end{equation}
We have maintained an $s$ dependence which comes from an ambiguity in defining at which low energy $s$ we evaluate $f$. In a weakly coupled theory $g_* \ll 1$ for which we are applying tree level positivity bounds, we must have $f$ positive for all $s$ in the range $4m^2-\Lambda^2<s<\Lambda^2$. We note that for all the scatterings involving the tensor modes, the contributions from the operator $c$ only arise at the scale $M$. This is despite the fact that in the action some of them would come at the $\Lambda_2$ scale. Indeed, very schematically, the contribution to the $TSTS$ channel would scale as
\be
\mathcal L_{TSTS}=c\Lambda_3^4 \left(\frac{\partial}{\Lambda_3}\right)^4\left(\frac{h}{M}\right)^2\left(\frac{\partial^2 \pi}{\Lambda_3^3}\right)^2=\frac{c}{M^2\Lambda_3^{6}}\partial^4 h^2(\partial ^2\pi)^2=\frac{c}{\Lambda_2^8}\partial^4 h^2(\partial ^2\pi)^2\,,
\ee
making clear that this is a $\Lambda_2$ scale operator. However, this contributes to the scattering amplitude with terms of the form $\Delta A_{TSTS} \sim c s^{4-n} m^{2n}/\Lambda_2^8$ which for $n=2$ gives a contribution to $f$ of the form $cm^4/\Lambda_2^8=c/M^4$ as we see above. \\

From the above expressions we see that the contributions from the operator \eqref{c_op} can make these bounds positive when the leading contribution is zero (\emph{i.e.} $\lambda_1=\lambda_3=0$). Moreover, as explained in Appendix \ref{app:indef}, this operator is allowed by all possible indefinite polarization bounds. It is then important to notice that the remaining definite helicity bounds also involve the quartic non-derivative operator $\lambda_4$ (or $\kappa_4^{(h)}$ in our conventions) leading to
\begin{align}
&f(VVVV)_-=-\frac{15\lambda_1^2+4\lambda_1\lambda_3-4\lambda_3^2+4\lambda_4}{16\Lambda_2^4}+\frac{c}{4M^4}>0\,,\nn \\
&f(VSVS)=-\frac{3\lambda_1^2-8\lambda_1\lambda_3-12\lambda_3^3+8\lambda_4}{48\Lambda_2^4}+\frac{c}{4M^4}>0\,,\\
&f(SSSS)=-\frac{5\lambda_1^2+6\lambda_1\lambda_3+\lambda_3^3+2\lambda_4}{9\Lambda_2^4}+\frac{\left(488 m^8-672 m^6 s+408 m^4 s^2-120 m^2 s^3+15 s^4\right)}{144 m^8M^4}c>0\,.\nn
\end{align}
Setting the cubic couplings to zero and demanding the above expressions to be positive then leads to the constraint $\lambda_4<0$. Hence we conclude that the positivity bounds applied on the EFT of a single pseudo-linear spin-2 field, as far as analysed in~\cite{Bonifacio:2016wcb} do not rule out the theory. In particular, the quartic operator $\varepsilon\varepsilon hhhh$ is still allowed by the analysis of~\cite{Bonifacio:2016wcb} \emph{i.e.} we could still have a non-zero negative $\lambda_4$.  However, upon pushing the positivity bounds further and considering indefinite polarization scattering, setting $\alpha_{S}=0$,  $\alpha_{V1}=0$ (see next section) and $\lambda_1=\lambda_3=0$ we further get the following bound:
\begin{equation}
    f=-\frac{\alpha_{V2}^2 \left(2 \beta_{S} \left(\beta_{S}-\sqrt{3} \beta_{T1}\right)+3 \beta_{V1}^2\right)}{12 m^2 M^2 x^2}\lambda_4\geq0.
\end{equation}
This inequality implies  $\lambda_4=0$ as the numerator  $\alpha_{V2}^2 \left(2 \beta_{V2} \left(\beta_{V2}-\sqrt{3} \beta_{T1}\right)+3 \beta_{V1}^2\right)$  can be both positive and negative for different choices of $\beta$'s and so all the leading terms are forced to be zero.  Again, while the couplings $\lambda_1$, $\lambda_3$ and $\lambda_4$ do need to vanish in order for the positivity bounds to be marginally satisfied, we have demonstrated that higher order operators naturally arising in the EFT can have non-zero positive contributions to $f$ and this provides a small technical window for the theory to live in, albeit one that is far less interesting than imagined at the outset.

\section{Bounds for Pseudo Linear Interactions}\label{sec:bounds}

\subsection{$hh\rightarrow hh$ Scattering}\label{hhtohh}

Now we apply the forward limit positivity bounds to a theory of two interacting pseudo-linear spin-2 fields. By considering some particular choices of polarizations of the ingoing and outgoing particles we shall show that the theory of two interacting pseudo-linear spin-2 fields can be ruled out by positivity bounds. For this it will be enough to consider the $hh\to hh$ (or, equivalently, $ff\to ff$) scatterings. We note that this scattering channel is independent on the ratio $\gamma$ of the interaction scales.

\subsubsection*{Definite Transversity States}

From the $hh\rightarrow hh$ definite transversity-2 scattering amplitude ($\alpha_{+2}=\beta_{-2}=1$, $\alpha_{-2}=\beta_{+2}=\alpha_{\pm1}=\beta_{\pm1}=\alpha_{0}=\beta_{0}=0$) we get
\begin{align}
\begin{split}
\label{eq:ineq1}
   f=&-\frac{1}{768 m^2 M^2x^8} \bigg[16 x^4 \left(5 c_{1} (a_{3}+2a_{6})+(a_{3}+2a_{6})^2-2 c_{1}^2\right)+8 c_{1} x^2 (a_{3}+2a_{6}+c_{1})+4 c_{1}^2\bigg.\\
   &+\bigg.3 x^6 \left(336a_{1}^2+420a_{1} \kappa_3^{(h)}+32 \left(2 c_{1} (a_{3}+2a_{6})+(a_{3}+2a_{6})^2+4 c_{1}^2\right)+273 \kappa_3^{(h)2}\right)\bigg]\geq 0\,.
\end{split}
\end{align}
A key point is that this definite helicity amplitude does not receive a contribution the quartic contact term $\kappa_4^{(h)}$. This expression can be written in the following form:
\begin{equation}
\label{eq:matrix eq}
    f=v^T \hat M v\ge 0\,,
\end{equation}
where $v=(\kappa_3^{(h)},a_{1},c_{1},a_{3},a_{6})$ and

\begin{equation}\label{eq:hessian}
   \hat M= \frac{1}{M^2}\left(
\begin{array}{ccccc}
-\frac{273}{128 m^2 x^2} & -\frac{105}{64 m^2 x^2} &0 & 0 & 0 \\
 -\frac{105}{64 m^2 x^2} & -\frac{21}{8 m^2 x^2} & 0 & 0 &0 \\
 0 & 0 &-\frac{96 x^6-8 x^4+2 x^2+1}{96 m^2 x^8} &-\frac{24 x^4+10 x^2+1}{96 m^2 x^6} & -\frac{24 x^4+10 x^2+1}{48 m^2 x^6} \\
 0 &0 & -\frac{24 x^4+10 x^2+1}{96 m^2 x^6} & -\frac{6 x^2+1}{24 m^2 x^4} &-\frac{6 x^2+1}{12 m^2 x^4} \\
 0 &0 &-\frac{24 x^4+10 x^2+1}{48 m^2 x^6} &-\frac{6 x^2+1}{12 m^2 x^4} &-\frac{6 x^2+1}{6 m^2 x^4}\\
\end{array}
\right).
\end{equation}
The matrix $\hat M$ consists of two block matrices that can be defined as
\be\label{AB}
\begin{split}
&A_{2\times 2}=-\frac{3}{128m^2M^2x^2}
\begin{pmatrix}
91&70\\
70&112
\end{pmatrix}\,,\\
&B_{3\times 3}=-\frac{(6x^2+1)}{96m^2M^2x^6}
\begin{pmatrix}
\frac{1}{x^2}(16x^4-4x^2+1)&4x^2+1&2(4x^2+1)\\
4x^2+1&4&8\\
2(4x^2+1)&2&4
\end{pmatrix}\,.
\end{split}
\ee
The eigenvalues of $A_{2\times 2}$ are
\be
\lambda_{A\pm}=\frac{-21}{256m^2M^2x^2}\left(29\pm \sqrt{409}\right)<0\,,
\ee
and are always negative. This means that the $(\kappa^{(h)}_3, a_1)$ subset in $v$ can only satisfy the inequality \eqref{eq:matrix eq} if both $\kappa_3^{(h)}=a_{1}=0$.

The matrix $B_{3\times 3}$ is apparently degenerate since the last two rows only differ by a factor of two giving one zero eigenvalue. This means that the two interaction terms, $a_3\varepsilon\varepsilon (\partial^2h)hf$ and $a_6\varepsilon\varepsilon (\partial^2 f)hh$, give equivalent contributions to the scattering between two transversity-2 states. This is apparent from the scattering amplitude \eqref{eq:ineq1} since $a_3,a_6$ only appear in the combination $a_3+2a_6$. It can also be seen when deriving the expression for the interaction vertex for an ingoing transversity-2 state. The remaining two non-zero eigenvalues are in turn
\begin{align}
\lambda_{B\pm}=-\frac{\left(6 x^2+1\right) }{192 m^2 M^2x^8}\left(36 x^4-4 x^2+1\pm\sqrt{336 x^8+192 x^6+28 x^4-8 x^2+1}\right)\leq 0\,.
\end{align}
One can check that these eigenvalues are both non-positive. Moreover, there is no value of $x$ for which both eigenvalues $\lambda_{B\pm}$  vanish simultaneously.
We also see that for $x=0$ and $x=\pm \frac{1}{2}$ at least one of the eigenvalues, $\lambda_{B-}$, vanishes. The case of $x=m_1/m_2=0$ corresponds to the situation when one of the two fields has no non-derivative self-interactions. However, in this limit $f$ is infinite, so we will not consider it any further. The case when $x=\pm \frac{1}{2}$ we shall explore in more detail below. \\

The total set of the eigenvalues is thus $\{\lambda_{A\pm}, \lambda_B=0, \lambda_{B\pm}\} $ and $M$ is negative semi-definite. Therefore, only the equality sign is allowed in \eqref{eq:matrix eq}, \emph{i.e.} the quantity $f$ in \eqref{eq:ineq1} is never strictly positive. As was said before, for $f$ to vanish it is necessary to demand that $\kappa_3^{(h)}=a_{1}=0$. Then, the remaining three quantities that determine the positivity (or non-negativity, to be more precise) are $\tilde a_3\equiv a_3+2a_6, c_1, x$. The quantity $f$ simplifies to
\be
f=-\frac{(6x^2+1)}{192 m^2 M^2x^8}\left[4x^2\tilde a_3^2+2c_1x^2(4x^2+1)\tilde a_3+c_1^2(16x^4-4x^2+1)\right]\,.
\ee
Demanding that it vanishes can be viewed as solving a quadratic equation for $\tilde a_3$ in terms of the other parameters $c_1$ and $x$. The discriminant of that equation can be found to be $D=-12c_1^2x^2(4x^2-1)^2\leq 0$. Thus, the equation $f=0$ can only have one real root in the case when the discriminant is zero. This can happen in two cases: $(i)$ when $c_1=0$ we find that $\tilde a_3 = 0$ for any value of $x$; $(ii)$ when $x=\pm\frac{1}{2}$ we find that $\tilde a_3=-2c_1$. The latter case is exactly reflecting the situation when one of the two eigenvalues is vanishing. We emphasize again that our results here can only constrain the combination of couplings $\tilde a_3\equiv a_3+2a_6$ thus leaving some freedom in the full parameter space even if we require that $\tilde a_3=0$ or $\tilde a_3=-2c_1$.

In the next subsection we show that the two possibilities $(i)$ and $(ii)$ are ruled out by choosing the ingoing and outgoing particles in the states with SVT polarizations given in Table~\ref{tab:pseudo gen}.

\subsubsection*{Polarizations in SVT basis}

\begin{table}[H]\normalsize
    \centering
    \begin{tabular}{|c|c|c|c|c|}
    \hline
         $\alpha_{T1}$ $\beta_{T1}$&$\alpha_{T2}$ $\beta_{T2}$ & $\alpha_{V1}$ $\beta_{V1}$ & $\alpha_{V2}$ $\beta_{V2}$ & $\alpha_{S}$ $\beta_{S}$ \\
         \hhline{|=|=|=|=|=|}
         $-0.460,-0.140$ &$ -0.212,0.0655$ & $ 0.517,-0.819$ & $ 0.113,0.180$ & $0.680,0.523$ \\
         \hline
         $-0.377,-0.143$ &$0.248,0.208$& $0.789,0.174$ &$-0.416,-0.313$&$0.008,-0.
         899$\\
    \hline
    $1,0$ &$ 0,0$ & $ 0,0$ & $ 0,0$ & $0,1$ \\
         \hline
    \end{tabular}
    \caption{Special configurations of polarizations for $hh\rightarrow hh$ scattering that rule out the equality case in \eqref{eq:ineq1}.}
    \label{tab:pseudo gen}
\end{table}

Here we consider four specific choices of polarizations in SVT basis, some of them given in Table~\ref{tab:pseudo gen}:
\begin{itemize}
\item{}
The two positivity quantities $f_1$ and $f_2$ of the amplitudes obtained from the scattering of particles in states determined by the first two sets of polarizations in Table~\ref{tab:pseudo gen} can be multiplied by positive numbers $b_1$ and $b_2$ and added together in such a way that the coupling $\kappa_{4}^{(h)}$ cancels from the sum of the two amplitudes. Since each of them has to satisfy the positivity bound on its own and in the sum they have only been multiplied by positive numbers we can require that the sum of the two must itself be non-negative. The resulting inequality is thus
\begin{align}
\label{eq:ineq3}
&0\leq b_1f_1+b_2f_2=\frac{1}{m^2M^2x^8}\Bigg[a_6 c_1 \left(2x^8-5.27x^6+6.1x^4-0.865
x^2+0.142\right)\Bigg.\\
&\left.+a_6^2 x^2 \left(x^6-5.23x^4+0.082
x^2-0.008\right)+
\frac{c_1^2 \left(x^{10}-0.04
x^8-88.4 x^6+33.6 x^4+0.51x^2-1.08\right)}{x^2}\right]\,,\nn
\end{align}
which can be expressed in the form of \eqref{eq:matrix eq} with $v=(c_{1},a_{6})$. It can then be shown that $\hat M$ is negative definite for $x<2.16$. Therefore, the resulting inequality can be true only for
\be\label{region_x_3}
x\ge 2.16\,.
\ee
This already rules out the case $(ii)$ where one had to require that $x=\pm\frac{1}{2}$ leaving $\kappa_3^{(h)}=a_1=a_3+2a_6=c_1=0$ as the only possibility to (marginally) satisfy the positivity bounds.

\item{}
We obtain even more stringent constraints from scattering particles with the last set of polarizations in Table~\ref{tab:pseudo gen}. This gives the following inequality:
\begin{equation}
\label{eq:ineq2}
f=    -\frac{a_6^2 (8x^2-1)(2x^2-1)+4 a_6 c_1 \left(8 x^4-10 x^2+1\right)+4 c_1^2 \left(4 x^4-6 x^2+1\right)}{12 x^4 m^2M^2}\ge 0\,.
\end{equation}
This again can be treated as a quadratic equation for $a_6$ depending on $c_1$ and $x$. To determine the parameter regions when the expression above is non-negative, we first find its roots, \emph{i.e.} the values of $a_6$ when the scattering amplitude is zero. We find that the discriminant of the above equation (we drop the overall factor $1/6x^4$) is $D=-64c_1^2x^2(6x^4-9x^2+1)$ and can only be non-negative in the region
\be\label{region_x_1}
\frac{1}{12} \left(9-\sqrt{57}\right)\leq x^2\leq \frac{1}{12} \left(9+\sqrt{57}\right)\,,
\ee
(in numerical values $0.121\leq x^2\leq 1.379$). If we are now to combine this with the constraint \eqref{region_x_3} found above, we see that we are forced to be in the region where the discriminant is negative. This implies that $f$ as a function of $a_6$ is a parabola that does not cross the zero for any real value of $a_6$. Thus the amplitude can be positive only if the `$a_6$-parabola' lies in the upper half-plane. This for general values of $x$ can be achieved by demanding that the coefficient of $a_6^2$ in $f$ is positive which happens if $-(8x^2-1)(2x^2-1)>0$ giving
\be\label{region_x_2}
\frac{1}{8}\leq x^2\leq\frac{1}{2}\,.
\ee
This is incompatible with $x\geq 2.16$ and thus we conclude that the two inequalities \eqref{eq:ineq3} and \eqref{eq:ineq2} can only be true simultaneously if both couplings $c_1, a_6$ vanish. This breaks the degeneracy in $a_3$ and $a_6$ that we saw in the previous subsection where we were only able to constrain $a_3+2a_6=0$. The analysis presented here thus allows us to conclude that all the couplings considered so far must vanish:
\be
\kappa_3^{(h)}=a_1=a_3=a_6=c_1=0\,.
\ee

\item{}
Finally, let us constrain the coupling $\kappa_4^{(h)}$ in the case when all the other couplings are vanishing, as required by the positivity bounds.
We find that this case is ruled out since by setting $\alpha_{S}=0$ and $\alpha_{V1}=0$. By leaving all the other polarizations arbitrary we get the following bound:
\begin{equation}
    f=-\frac{\alpha_{V2}^2 \left(2 \beta_{S} \left(\beta_{S}-\sqrt{3} \beta_{T1}\right)+3 \beta_{V1}^2\right)}{2 m^2 M^2 x^2}\kappa_4^{(h)}\geq0.
\end{equation}

This inequality implies  $\kappa_4^{(h)}=0$ as the numerator  $\alpha_{V2}^2 \left(2 \beta_{V2} \left(\beta_{V2}-\sqrt{3} \beta_{T1}\right)+3 \beta_{V1}^2\right)$  can be both positive and negative for different choices of $\beta$'s.
\end{itemize}

Hence, we conclude there is no allowed region of parameter space for an EFT including two interacting pseudo-linear spin-2 field consistent with positivity bounds. As we will see in section \ref{gen}, this conclusion for this particular scattering process can be generalized to multiple interacting pseudo-linear spin-2 fields.

\subsection{$ff\rightarrow ff$ and $hf\to hf$ Scattering}

The scattering amplitude of $ff\rightarrow ff$ can be obtained from the $hh\rightarrow hh$ scattering amplitude, up to an overall factor of mass ratio, by changing $m_1\rightarrow m_2$, the non-derivative mixing couplings $c_1\rightarrow \frac{m^2_{2}}{m^2_{1}}c_2$, $a_1\rightarrow a_2$, $a_3\rightarrow a_4$, $a_6\rightarrow a_5$ and $\kappa_{3,4}^{(h)}\rightarrow \kappa_{3,4}^{(f)}$. Hence, we can recover the previous conclusion for the operators contributing to the $ff\rightarrow ff$, \emph{i.e.} that
\be
\kappa_3^{(f)}=\kappa_4^{(f)}=a_2=a_4=a_5=c_2=0\,,
\ee
 and there is no region of parameter space compatible with the positivity bounds.

 Since there are no cubic couplings, the quartic operator, $\lambda\varepsilon \varepsilon hhff$, contributing to the $hf\rightarrow hf$ scattering amplitude is ruled out as mentioned in \cite{alberte2020positivity}.

\section{Extension to Any Number of Massive Pseudo-Linear Spin-2 Fields}\label{gen}

In this section we consider multiple massive pseudo-linear spin-2 fields. We add additional fields, $f^{(i)}$ with $i=2,...,N$ and $N$ denoting the total number of spin-2 fields with mass $m_i$. The interacting terms contributing to the $hh\rightarrow hh$ scattering amplitude are:
\be\label{model1}
\begin{split}
 g_*^2 \L^{(i)}=
&\frac{\M_1\M_i}{4}\left(2a^{(i)}_3\varepsilon\varepsilon (\partial^2h)hf^{(i)}+2a^{(i)}_6\varepsilon\varepsilon (\partial^2 f^{(i)})hh\right)+\frac{m_i^2\M_1\M_i}{4}2c^{(i)}_1\varepsilon\varepsilon I hhf^{(i)}+...\,.
\end{split}
\ee
By choosing the  polarizations of the previous section, \eqref{eq:matrix eq} can be generalised. Since there are no mixing terms in the $hh\rightarrow hh$ amplitude between $h$ self-interaction couplings and the $h-f^{(i)}$ couplings, schematically, $\hat M$ and $v$ in \eqref{eq:matrix eq} take the following form:

\begin{equation}\label{eq:hessian2}
   \hat M= \left(
   \setlength{\arraycolsep}{0pt}
\begin{array}{ccccc}
 \fbox{$A_{2\times2}$} &0 & 0 & \dots &0  \\
0 &\fbox{$B^{(1)}_{3\times3}$} & 0 & \dots &0 \\
0 & 0 &\fbox{$B^{(2)}_{3\times3}$} &\dots & 0 \\
\vdots& \vdots &\vdots & \ddots  & \vdots \\
0 & 0 & 0  &\dots & \fbox{$B^{(i)}_{3\times3}$}\\
\end{array}
\right), v=(\kappa_3^{(h)},a_{1},c^{(1)}_{1},a^{(1)}_{3},a^{(1)}_{6},... ,c^{(i)}_{1},a^{(i)}_{3},a^{(i)}_{6}).
\end{equation}
Where the matrices $B^{(i)}_{3\times3}$ are all in the form of the $3\times3$ block diagonal matrix in \eqref{eq:hessian} and \eqref{AB}. Thus, in \eqref{eq:hessian2} all the block matrices are negative semi-definite with $x$ substituted by $x^{(i)}=\frac{m_{(i)}}{m_2}$ for $i=1,...,N$. Hence, the previous analysis can be easily extended to this case and so the coefficients of the leading operators contributing to the tree-level $hh\rightarrow hh$ scattering amplitude must be zero for a theory with any number of pseudo-linear massive spin-2 fields.

\section{Discussion}\label{sec:conclusions}

In this article we have extended the discussion of positivity bound constraints on effective field theories of multiple spin-2 particles in \cite{alberte2020positivity} to the case of pseudo-linear spin-2 theories where the symmetries broken by the mass term are linear diffeomorphisms/spin-2 gauge invariance. By applying forward limit positivity bounds we found that non-zero operators in the action \eqref{model0} (except for $d_1, d_2$) lead to a theory which does not have a local, Lorentz invariant, causal and unitary UV completion. Our analysis for the operators contributing to $hh\rightarrow hh$ tree-level scattering works for any number of pseudo-linear massive spin-2 fields. This is consistent with previous work excluding the possibility of a single pseudo-linear spin-2 field.
However, all of these statements should be understood within the context of an effective field theory expansion. Technically speaking it is possible to satisfy the leading forward limit positivity bounds by having the leading interactions zero (or parametrically smaller), and then using higher derivative operators in the EFT to satisfy positivity. Thus the more accurate statement is that if the leading operators are marginally ruled out, then it means that they must be suppressed in such away that higher derivative operators contribute equally or dominantly to the desired bounds. This nevertheless has a profound effect on the assumed structure of the effective field theory expansion, and it is quite possible that the application of more general positivity bounds, such as for example the non-forward limit bounds \cite{deRham:2017avq,deRham:2017zjm} exclude the EFTs entirely. \\

Our particular results extend easily to any number of pseudo-linear spin-2 fields. A similar (but not identical in origin) result that adding more fields does not increase the allowed region in the parameter space was also seen in \cite{alberte2020positivity} where it was observed that adding an extra massive spin-2 field to ghost-free massive gravity shrinks the allowed region for self-couplings of the other field. This is due to a combination of the increase in the number of constraints with increasing number of fields and the knowledge of an extra pole to subtract automatically strengthening any low energy bounds. In the case of two spin-2, constraining $d_1$ and $d_2$ operators would required to go beyond 2--2 elastic scattering amplitudes for which would require an extension of the standard positivity bounds formalism. Similarly in the case of multiple spin-2 there are interactions which do not contribute to $h^{(i)}h^{(j)} \rightarrow h^{(i)}h^{(j)} $ that we have not excluded. \\

Overall these results yet again demonstrate the power of the application of positivity bounds to effective field theories, and in particular those of spin-2. We stress again that these bounds are all derived based on a standard local Lorentz invariant UV completion and that giving up any one of these assumptions can lead to different conclusions. For instance the spin-2 states that arise in the condensed matter context are not constrained by these requirements. In the relativistic case it is also possible that the assumption of locality is not appropriate, particular to spin-2 states arising in an underlying gravitational theory.

\bigskip
\noindent{\textbf{Acknowledgments:}}
AJT and CdR would like thank the Perimeter Institute for Theoretical Physics for its hospitality during part of this work and for support from the Simons Emmy Noether program.
LA is supported by the European Research Council under the European Union's Seventh Framework Programme (FP7/2007-2013),
ERC Grant agreement ADG 339140.
The work of AJT and CdR is supported by an STFC grant ST/P000762/1. JR is supported by an STFC studentship. CdR thanks the Royal Society for support at ICL through a Wolfson Research Merit Award. CdR is supported by the European Union's Horizon 2020 Research Council grant 724659 MassiveCosmo ERC-2016-COG and by a Simons Foundation award ID 555326 under the Simons Foundation's Origins of the Universe initiative, `\textit{Cosmology Beyond Einstein's Theory}'. AJT thanks the Royal Society for support at ICL through a Wolfson Research Merit Award.

\bigskip

\appendix

\section{Polarization Tensors}\label{sec:transv}
In this section we give  the polarization tensors both in the transversity basis and in the SVT basis. In our analysis we mainly use the SVT basis, however, part of the $hh\to hh$ analysis of Subsection \ref{hhtohh} was done in the transversity basis. Since the latter is less common in the literature we present it in more detail. This basis was first used in the modern context in \cite{deRham:2017zjm} and was shown to be a convenient basis for positivity bounds away from the forward limit.

Throughout this work we use a frame where the momenta of ingoing and outgoing particles are parametrized as
\be
p^\mu_i=\left( E_i,p\sin\theta_i,0,p\cos\theta_i\right)\,,
\ee
where $i = 1,\dots,4$, the angles are $\theta_1=0\,,\theta_2=\pi$, $\theta _3=\theta\,,\theta_4=\pi+\theta$ and the energies satisfy $ E_1=E_3$, $E_2=E_4$. We only consider the forward limit ($\theta=0$) scattering amplitudes. For more detailed conventions we refer to the Appendix~B of \cite{alberte2020positivity}.

\subsection{Transversity Basis}

The polarization vectors in the transversity basis in the frame where the momentum of the corresponding particle is $p^\mu=(E,0,0,p)$ are defined as follows \cite{deRham:2017zjm}:
\begin{align}
   &\epsilon^{\mu}_{\tau=\pm 1} = \frac{i}{\sqrt{2}m}(p,\pm im,0,E)\,, \\
    &\epsilon^{\mu}_{\tau=0} = (0,0,1,0)\,.
\end{align}
To find the polarization tensors satisfying $p_\mu\epsilon^{\mu\nu}_\tau=0$, $\epsilon_\tau^\mu\,_\mu=0$ we express them as a combination of the polarization vectors with the appropriate Clebsch-Gordan coefficients:
\begin{align}
    &\epsilon^{\mu\nu}_{\tau=\pm 2} = -\epsilon^{\mu}_\pm    \epsilon^{\nu}_\pm\,, \\
    &\epsilon^{\mu\nu}_{\tau=\pm 1} = \frac{1}{\sqrt{2}}(\epsilon^{\mu}_\pm\epsilon^{\nu}_0 +
   \epsilon^{\mu}_0\epsilon^{\nu}_\pm)\,, \\
  &\epsilon^{\mu\nu}_{\tau=0} = - \frac{1}{\sqrt{6}}
    (\epsilon^{\mu}_+\epsilon^{\nu}_-
    + \epsilon^{\mu}_-\epsilon^{\nu}_+
    + 2\epsilon^{\mu}_0\epsilon^{\nu}_0)\,.
\end{align}
Hence, the polarization tensors are
\begin{align}
   &{\epsilon^{\mu\nu}_{\tau=\pm 2}} =\frac{1}{2m^2}
    \begin{pmatrix}
        p^2     & \pm imp  & 0  & p E\\
       \pm im p  &-m^2   & 0 &\pm imE\\
        0 & 0 & 0 & 0\\
        p E  &  \pm imE & 0 &E^2
    \end{pmatrix}\,,\\
   & {\epsilon^{\mu\nu}_{\tau=\pm 1}} = \frac{1}{2m}
    \begin{pmatrix}
        0 & 0 & ip & 0 \\
        0 & 0 &\mp m & 0 \\
        ip &\mp m & 0 & i E \\
        0 & 0 & i E & 0
    \end{pmatrix}\,,\\
   & {\epsilon^{\mu\nu}_{\tau=0}} = \frac{1}{\sqrt{6}m^2}
    \begin{pmatrix}
        p^2 &0 & 0 & pE \\
      0 &m^2 & 0 &0\\
        0 & 0 & -2m^2 & 0 \\
        pE &0 & 0 & E^2
    \end{pmatrix}\,.
\end{align}
As in the SVT basis in \eqref{indefpolhel}, the entire configuration of transversities can be specified by ten (real) numbers, $\alpha_1,..,\alpha_5,\beta_1,...\beta_5$, as
\begin{align}
\label{indefpol_trans}
\begin{split}
&\epsilon^{(1)}=\alpha_1\epsilon_{+2}+\alpha_2\epsilon_{+1}+\alpha_3\epsilon_{0}+\alpha_4\epsilon_{-1}+\alpha_5\epsilon_{-2}\,, \\
&\epsilon^{(2)}=\beta_1\epsilon_{+2}+\beta_2\epsilon_{+1}+\beta_3\epsilon_{0}+\beta_4\epsilon_{-1}+\beta_5\epsilon_{-2}\,,\\
&\epsilon^{(3)}=\epsilon^{(1)}\,,\\
&\epsilon^{(4)}=\epsilon^{(2)}\,.
\end{split}
\end{align}

\subsection{SVT basis}
The coefficients in the indefinite transversity polarization states above are related to the $\alpha$'s and $\beta$'s in SVT basis in  \eqref{indefpolhel} by the following transformation:

\begin{equation}\label{SVT to Tran}
\left(
\begin{array}{ccccc}
\alpha_{T1}\\
\alpha_{T2}\\
\alpha_{V1}\\
\alpha_{V2}\\
\alpha_{S}\\
\end{array}
\right)=
\left(
\begin{array}{ccccc}
 -\frac{1}{2 \sqrt{2}} & 0 & \frac{\sqrt{3}}{2} & 0 & -\frac{1}{2 \sqrt{2}} \\
 0 & \frac{1}{\sqrt{2}} & 0 & -\frac{1}{\sqrt{2}} & 0 \\
 -\frac{1}{\sqrt{2}} & 0 & 0 & 0 & \frac{1}{\sqrt{2}} \\
 0 & \frac{1}{\sqrt{2}} & 0 & \frac{1}{\sqrt{2}} & 0 \\
 \frac{\sqrt{\frac{3}{2}}}{2} & 0 & \frac{1}{2} & 0 & \frac{\sqrt{\frac{3}{2}}}{2} \\
\end{array}
\right)
\left(
\begin{array}{ccccc}

\alpha_{-2}\\
\alpha_{-1}\\
\alpha_{0}\\
\alpha_{+1}\\
\alpha_{+2}\\
\end{array}
\right).
\end{equation}

\section{Indefinite Bounds for the Higher Operator}\label{app:indef}
We can conclude from Section \ref{sec:bounds} that the coefficients $a_1$, $\kappa_3^{(h)}$ and $\kappa_4^{(h)}$ in Eq. \eqref{action_single} must all be equal to zero. Then the bound from indefinite scattering is found to be:
\allowdisplaybreaks

\begin{align}
\begin{autobreak}
\frac{c}{144 m^8 M^4}
\bigg(4 m^8 \bigg(\alpha_S^2
\bigg(122 \beta _S^2+9
\bigg(\beta _{T1}^2
+\beta _{T2}^2
+\beta _{V1}^2
+\beta _{V2}^2\bigg)\bigg)
+6 \alpha _S \beta _S \bigg(15 \alpha _{T1} \beta _{T1}
-15\alpha _{T2} \beta _{T2}
-17 \alpha _{V1} \beta _{V1}
+17 \alpha _{V2}
\beta _{V2}\bigg)
+9 \bigg(\alpha _{T1}^2 \bigg(\beta _S^2
+2 \beta_{T1}^2
+\beta _{T2}^2
+\beta _{V1}^2
+\beta _{V2}^2\bigg)
+\alpha_{T2}^2 \bigg(\beta _S^2
+\beta _{T1}^2
+2 \beta _{T2}^2
+\beta _{V1}^2
+\beta _{V2}^2\bigg)
+\beta _S^2 \alpha _{V1}^2
+\beta _S^2 \alpha_{V2}^2
-2 \alpha _{T1} \beta _{T1} \bigg(\alpha _{T2} \beta _{T2}
+3 \alpha _{V1} \beta _{V1}
-3 \alpha _{V2} \beta _{V2}\bigg)
+\beta_{T1}^2 \alpha _{V1}^2
+\beta _{T1}^2 \alpha _{V2}^2
+\beta _{T2}^2
\alpha _{V1}^2
+6 \alpha _{T2} \beta _{T2} \bigg(\alpha _{V1} \beta_{V1}
-\alpha _{V2} \beta _{V2}\bigg)
+\beta _{T2}^2 \alpha _{V2}^2
+8\alpha _{V1}^2 \beta _{V1}^2
+\beta _{V1}^2 \alpha _{V2}^2
+\beta_{V2}^2 \bigg(\alpha _{V1}^2
+8 \alpha _{V2}^2\bigg)
-14 \alpha _{V1}
\beta _{V1} \alpha _{V2} \beta _{V2}\bigg)\bigg)
-12 m^6 s \bigg(56\alpha _S^2 \beta _S^2
+3 \alpha _S \beta _S \bigg(8 \alpha _{T1}
\beta _{T1}
-8 \alpha _{T2} \beta _{T2}
-19 \alpha _{V1} \beta _{V1}
+19\alpha _{V2} \beta _{V2}\bigg)
+9 \bigg(\alpha _{V1} \beta_{V1}
-\alpha _{V2} \beta _{V2}\bigg) \bigg(-\alpha _{T1} \beta _{T1}
+\alpha _{T2} \beta _{T2}
+2 \alpha _{V1} \beta _{V1}
-2 \alpha_{V2} \beta _{V2}\bigg)\bigg)
+6 m^4 s^2 \bigg(68 \alpha _S^2 \beta_S^2
+12 \alpha _S \beta _S \bigg(\alpha _{T1} \beta _{T1}
-\alpha_{T2} \beta _{T2}
-5 \alpha _{V1} \beta _{V1}
+5 \alpha _{V2} \beta_{V2}\bigg)
+9 \bigg(\alpha _{V1} \beta _{V1}
-\alpha _{V2} \beta_{V2}\bigg){}^2\bigg)
-60 m^2 s^3 \alpha _S \beta _S \bigg(2 \alpha
_S \beta_S-\alpha _{V1} \beta _{V1}
+\alpha _{V2} \beta _{V2}\bigg)
+15 s^4 \alpha _S^2 \beta _S^2\bigg)
\end{autobreak}
\end{align}
By numerically varying this equations with $\alpha$'s, $\beta$'s and $s$ we found the minimum to be  $0.114\frac{m^4c}{\Lambda^8_{2}}>0$, therefore this operator is allowed by all the indefinite bounds.

\bibliographystyle{JHEP}
\bibliography{references}

\end{document}